# THERMOPLASTIC POLYURETHANE-GRAPHENE NANOPLATELETS MICROCELLULAR FOAMS FOR ELECTROMAGNETIC INTERFERENCE SHIELDING

*Maria Giovanna Pastore Carbone[1*], Maxime Beaugendre[2], Can Koral[3], Anastasios C. Manikas[1,4], Nikolaos Koutroumanis[1], Gianpaolo Papari[5], Antonello Andreone[3,5], Ernesto Di Maio[6] and Costas Galiotis[1,2]*

[1]Foundation of Research and Technology Hellas, Institute of Chemical Engineering and High Temperature Processes, Stadiou St. Platani, GR-26504 Patras (Greece)
[2]ESPCI Paris, CNRS, PSL University, 10 Rue Vauquelin, 75005 Paris (France)
[3]INFN Naples Unit, I-80126, Naples (Italy)
[4]Department of Chemical Engineering, University of Patras, GR-26504 Patras (Greece)
[5]Department of Physics, University of Naples "Federico II", I-80125, Naples (Italy)
[6]Dipartimento di Ingegneria Chimica, dei Materiali e della Produzione Industriale, University of Naples "Federico II", I-80125, Naples (Italy)

*Corresponding author: mg.pastore@iceht.forth.gr

## Abstract

The incorporation of graphene-related materials as nanofiller can produce multifunctional foams with enhanced specific properties and density reduction. Herein we report on the preparation of microcellular thermoplastic polyurethane/graphene foams by batch foaming. Solution blending was first adopted to disperse graphene nanoplatelets (GNP) in the elastomeric matrix. Then a foaming process based on the use of supercritical $CO_2$ was adopted to produce the microcellular TPU/GNP composite foams with graphene content up to 1 wt%. The EMI shielding behaviour of the TPU/GNP foams has been assessed in the THz range, and has revealed their potential in comparison with other graphene-filled foams presented in literature, that exhibit similar specific shielding effectiveness but at much higher content of graphene-related materials (10-30 wt%).

# 1. Introduction

Electromagnetic pollution is becoming a serious problem due to the fast development of electronic and wireless technology. Therefore, over the last few years, the design and the fabrication of electromagnetic interference (EMI) shielding materials have been gaining increased attention in the academic and industrial fields[1]. Compared to the conventional EMI shielding materials (e.g. metals), polymer composites containing carbon-based nanofillers have the advantage of being lightweight, low-cost and easy-processable, corrosion resistant and broad bandwidth absorbing[2]. Among carbon-based materials, such as natural graphite flakes, expanded graphite (EG), graphene nanoplatelets (GNP), graphene oxide (GO), reduced graphene oxide (rGO), single and few-layer graphenes have been paid the most attention due to their outstanding physio-mechanical properties[3]. In particular, graphene, the perfect 2D crystal of covalently bonded carbon atoms, possesses a unique unusual combination of physical properties, such as remarkable mechanical, thermal and electrical properties, and represents the ideal candidate for the fabrication of novel materials with improved structural and functional properties, thus offering great promise for use in EMI shielding in special fields such as aerospace, weapon equipment, vehicles, and microelectronics[4].

Furthermore, the current challenge in EMI shielding for aerospace and electronics is to achieve high shielding effectiveness (SE) with low weight. Foaming represents an effective route to produce lightweight materials based on polymers and composites, and the preparation of polymer foams filled with carbon nanotubes or carbon nanofibers for EMI shielding has been widely reported[5, 6, 7]. In recent years, polymer foams containing graphene-related materials (GRM) have also been studied for EMI shielding due to the unique combination of superior electrical properties of the nanofiller and flexibility of the polymer matrix[8]. In particular, a great deal of effort has been put in the design and development of GRM-based polymer nanocomposites with specific arrangement of the filler into spatially-segregated 3D architectures (both bulk and porous), which can provide significant improvements in terms of structural and functional features[9].

Thermoplastic polyurethane (TPU) elastomers are a class of multi-block copolymers composed of soft and hard segments: the crystalline hard domains behave as physical crosslinking, while the amorphous soft domains provide typical rubber-like property. The final performance of the TPU can be easily tuned by controlling the ratio of soft and hard segments; due to their versatility, TPU elastomers have traditionally been used in automotive, electronics and construction[10]. The addition of GRM to TPU for development of nanocomposites with improved electrical properties[11, 12], sensing[13, 14] and EMI shielding applications[15] has been already reported.

Concerning lightweight applications, TPU with microcellular morphology have also been developed and are used in footwear, sports and leisure, clothing and medical applications[16]. Many methods have been proposed to create TPU foams, such as in situ polymerization using water as a foaming agent, gas foaming, salt leaching, phase inversion and water vapour induced phase separation [17]. Using supercritical fluids as blowing agent, and in particular carbon dioxide (sc-$CO_2$), has become a promising and efficient strategy for the preparation of microcellular polymeric foams[18]. The attractiveness of using sc-$CO_2$ is linked to the low cost, chemical stability, moderate critical conditions ($T_c$ =31.1 °C, $P_c$ =7.38 MPa), lowered burden on the environment compared to traditional blowing agents (e.g. HCFCs), and greater safety offered compared to flammable hydrocarbons[18]. This technology also possesses the great advantage of being easily scaled-up to industrial level. Recently, Chen et al. reported on the development of lightweight, electrically conductive TPU/graphene foams obtained via water vapour induced phase separation of nanocomposites previously produced by solution mixing, thus suggesting the potential of TPU nanocomposite foams for high-performance applications[17]. Furthermore, sonication-aided graphene impregnation technique and thermal induced phase separation have been proposed to produce conductive TPU foams for sensing applications[7, 19]. TPU/rGO composite foams have been also proposed for EMI shielding, and SE value of 21.8 dB in the X-band was achieved with only 3.17 vol% rGO loading owing to the multistage cellular structure with good conductive network[20].

In this study, we report a facile process to produce lightweight microcellular TPU/graphene foams for EMI shielding applications. Solution blending method was first adopted to fill TPU with graphene nanoparticles (GNP); afterwards, batch foaming process based on the use of sc-$CO_2$ was applied to prepare the microcellular TPU/GNP composite foams. Finally, the EMI shielding effectiveness of both the non-foamed and foamed composites was explored in the THz range.

## 2. Experimental Section

### 2.1. Materials and methods

*Materials*. A polyester-based thermoplastic polyurethane (Elastollan 890AN), gently supplied by BASF SE (Ludwigshafen, Germany), was used as polymer matrix. This TPU has a mass density of 1.22 g/cm$^3$, a hardness of 92 shore A and a melt flow index of 15-30 g/10 min (ASTM D1238). Graphene nanoplatelets (Elicarb® Graphene powder materials grade), with typical lateral size ca. 0.5-2 μm, was kindly supplied by Thomas Swan (Consett, UK).

*Preparation of TPU/GNP nanocomposites*. TPU/GNP composites with loadings ranging from 0 to 1 percent by weight (wt.%) were produced by solution blending method. TPU (2 g) was firstly dissolved in DMF (30 ml) at 80 °C for 5 hours, to give a 10 wt.% solution. Graphene nanoplatelets were dispersed in DMF with the aid of bath sonication for 2 hours. The amounts of graphene were calculated in order to obtain composites with 0.1 and 1 % wt. which corresponds to 0.0020 g and 0.0200 g respectively. Then the GNP dispersion was mixed to the polymer solution and stirred for 15 min and finally sonicated for 5 min. The mixture was then poured into a Petri dish where the solvent was left to evaporate at 60 $^{o}$C for 24 h and then dried under vacuum at 70 $^{o}$C for 1 day. For comparison, membranes of the neat polymer were fabricated in the same method without the addition of graphene.

*Preparation of TPU/GNP nanocomposite microcellular foams*. All foams were prepared by a batch foaming process with the aid of sc-$CO_2$. The foaming equipment utilized in this study is a custom-made pressure vessel with accurate control of the foaming processing variables[21]. The control of the temperature was achieved by means of a PID controller and a Pt100 sensor. A pressure transducer was used to measure pressure and the pressure history was registered by using a data acquisition system. A syringe pump 500D (Teledyne Isco, Lincoln, NE, USA) was used to define the pressure history, controlled by a personal computer via a RS232 connection. The pressure release system consists of a discharge valve and a pneumatic electro-valve. In a typical experiment, a rectangular sample, 0.1-0.2 mm in thickness and 10-15 mm in side was placed in the pressure vessel. After sample loading, the vessel was heated to the saturation temperature and pressurization started according to the defined program. At the end of the pressure program, pressure was rapidly released and samples were then removed from the vessel.

*Characterization*. The quality of graphene incorporated in the TPU matrix has been evaluated by using Raman spectroscopy. Raman spectra were acquired with the Micro-Raman spectrograph (InVia Reflex, Renishaw, UK) at 785 nm and with a100x objective; the laser power was kept at 1.2 mW on the sample to avoid laser-induced local heating. The polarization of the incident light was kept parallel to the applied strain axis. All Raman spectra were background-corrected.

Mechanical testing of the nanocomposites was performed by tensile experiments in an MTS R58 Mini Bionix machine with a strain rate of 100% s$^{-1}$. The produced materials were cut in strips of dimensions ~45 x 4 mm and, for each material type, five samples were tested to extract stress-strain curves. The Young's modulus was estimated through a linear regression analysis of the initial linear portion of the stress–strain curves.

The microstructure of the foams was observed with a scanning electron microscope (SEM) (S440, LEICA). The samples were first sectioned with a razor blade and then coated with gold using a sputter coater before SEM observation. Image analysis on the SEM micrograph was conducted by using the software ImageJ in order to estimate the average cell size. The density of the samples was calculated by measuring the weight and dividing it by the volume determined using the water displacement method, according to ASTM D792, with an analytical balance (Mettler Toledo, Columbus, OH).

The EMI shielding response of both the nanocomposites and the microcellular foams was investigated by using a femtosecond laser driven time domain spectrometer (Menlo Systems/ TeraK15). The system exploits fibre-coupled photoconductive switches both for THz emission and detection and is widely used for the characterisation of carbon loaded polymeric composites[22] and other advanced materials[23] in the frequency range 0.1 – 2 THz. Data acquisition was realized by means of a lock-in amplifier coupled with electronics and computer software. A standard setup with four polymethypenetene (TPX) lenses was used to collimate and focus the broad THz beam first impinging onto the sample plane and then transmitted to the detector. The transient electric field signal versus time was measured separately upon transmitting through the sample ($\hat{E}_{smpl}$) and through the free space used as reference ($\hat{E}_{ref}$). Time domain signals were converted into the frequency domain by applying a Fast Fourier Transform (FFT), and the complex transfer function, $\tilde{T}(\omega) = \frac{\hat{E}_{smpl}(\omega)}{\hat{E}_{ref}(\omega)}$, was measured with a resolution better than 5 GHz[23].

## 3. Results and Discussion

Raman spectroscopy was adopted to assess the quality of graphene in the nanocomposites produced by solution mixing (Figure 1). For neat TPU, the characteristic peak at 2930 cm$^{-1}$ is attributed to the stretching vibrations of –CH$_2$; the two weak peaks at 1730 and at 1698 cm$^{-1}$ are related to the stretching vibration of the carbonyl group, respectively, in the free configuration and in the hydrogen bond with nitrogen; the strong peak at 1615 cm$^{-1}$ corresponds to the aromatic breathing mode symmetric stretch vibration of C=C; the peak at 1538 cm$^{-1}$ is due to the C=C of urethane amide and the peak at 1445 cm$^{-1}$ is assigned to the bending vibrations of –CH$_2$ [24]. In the TPU/GNP nanocomposite, typical features of multi-layers graphene appear in the Raman spectrum, specifically the strong G peak at 1581 cm$^{-1}$, the 2D peak at 2646 cm$^{-1}$ and the D band, which appears as a shoulder at 1365 cm$^{-1}$ [25]. In Figure 2, the stress-strain curves from tensile experiments of the produced TPU/GNP nanocomposites are shown. All TPU samples show elastic behaviour up to a ~15% strain and the presence of GNP, even at small amounts, affects the polymer properties. More specifically, Young's modulus is increased from 42 MPa for neat polymer to 49 and 62 MPa (47.6% increase) by the addition of 0.1% and 1% GNP, respectively. The slight decrease of stress recorded at around 300% strain can be ascribed to possible grip slippage at high deformations and is not representative of the original properties of the material. However, it is important to note that the ultimate tensile stress is increased about 25% for both nanocomposites and that the maximum deformation remains as high as for the neat elastomer.

These results provide good evidence for the successful dispersion of graphene flakes inside the matrix and reveal that if an appropriate dispersion has been achieved, then the matrix properties can significantly be improved even with a negligible amount of graphene.

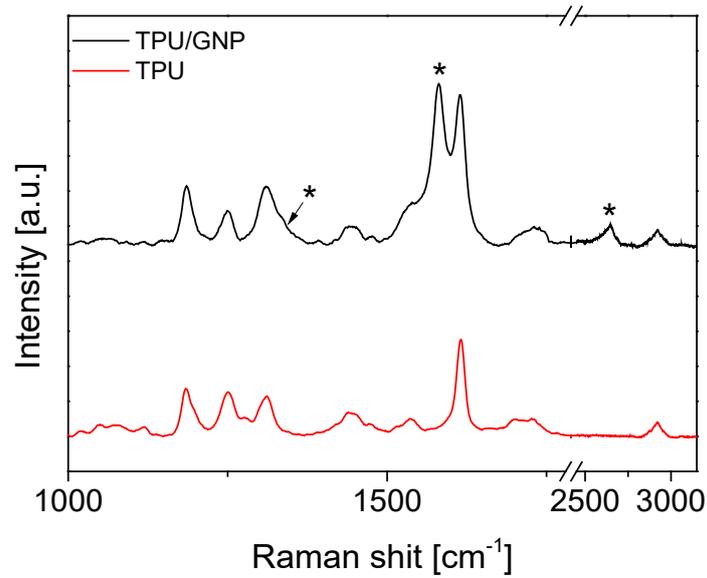

**Figure 1.** Raman spectra of neat TPU and of TPU/GNP nanocomposite. Spectroscopic features of multi-layers graphene are marked by asterisks.

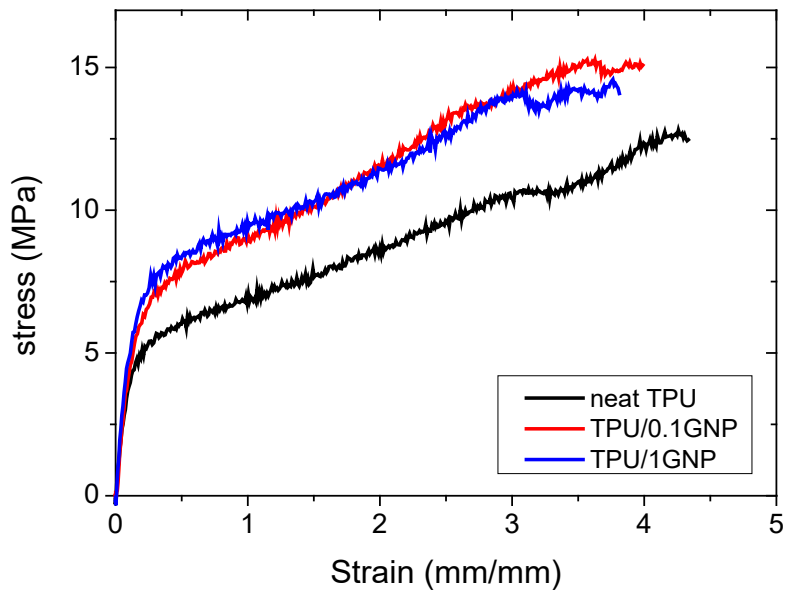

**Figure 2.** Representative stress-strain curves for neat TPU and its nanocomposites with GNP loadings of 0.1 and 1 wt%

Preliminary foaming tests have been performed in order to adjust saturation and foaming conditions. Figure 3a-b shows a typical SEM micrograph of the cross-section of the microcellular foam produced 100 °C and 240 bars. It is evident that the microcellular cells with an average size of ~10 μm were distributed throughout the foam. A typical optical micrograph of TPU/GNP nanocomposite foam with graphene loading of 1 wt% is shown in Figure 3c. The thickness of foam sheet was ~0.3-0.5 mm, and it was quite flexible under bending. Table 1 shows the density of TPU/GNP nanocomposite foams as a function of graphene content. The density of TPU slightly increased from 1.22 to 1.23 g/cm$^3$ upon the addition of GNP. After expansion, the density of TPU foam was 0.69 g/cm$^3$ and it is interesting to find that the introduction of graphene did not change substantially the density of nanocomposite foams.

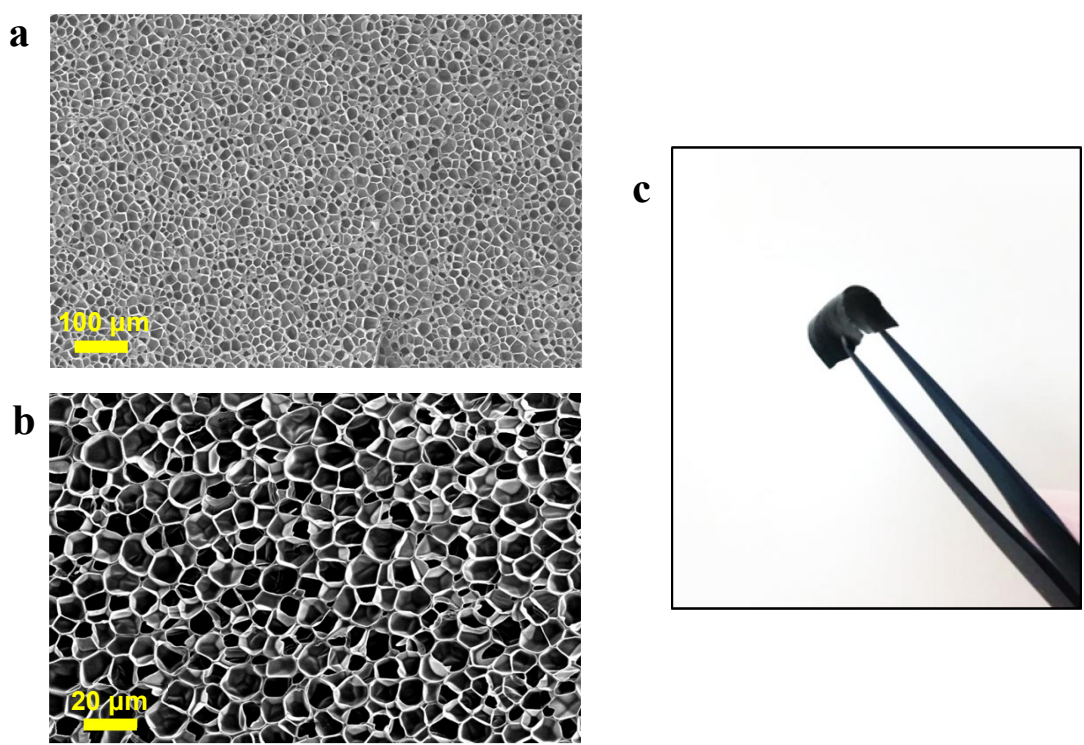

**Figure 3**. Representative cross-sectional SEM images (a,b) and photograph (c) of TPU/GNP foam.

| Graphene content in bulk (wt%) | Graphene content in bulk (vol%) | Bulk Density (g/cm³) | Graphene content in foam (vol%) | Foam Density (g/cm³) |
|---|---|---|---|---|
| 0 | 0 | 1.22 | 0 | 0.69 |
| 0.1 | 0.05 | 1.225 | 0.03 | 0.66 |
| 1 | 0.54 | 1.237 | 0.31 | 0.71 |

**Table 1**. Density Values of TPU/GNP nanocomposites (bulk) and foams with different graphene loadings.

In this study, we assessed the shielding properties of the TPU/GNP nanocomposites and microcellular foams in the range 0.3-1.0 THz, across almost one frequency decade. Terahertz shielding is gaining increasing attention since there has been recently a significant advance in the development of very high frequency electronics and devices for different applications, such as wireless communication, imaging, and sensing. Recently, graphene-related materials and their composites have been demonstrated to provide effective EMI shielding in the THz frequency domain[26, 27, 28, 29], mainly owning to their electro-conductive properties, which are responsible for reflecting and absorbing THz electromagnetic waves[30].

The EMI shielding efficiency (EMI SE) is measured in terms of the total attenuation in the incident electric field upon transmission through the barrier media (the shield). The total shielding efficiency ($SE_{TOT}$) in decibel (dB) units can be directly calculated from the electric field ratios of incident and transmitted electromagnetic waves:

$$SE_{TOT} = -20 \log(T) \qquad \text{eq. 1}$$

where

$$T = \left|\frac{\hat{E}_{smpl}}{\hat{E}_{ref}}\right| \qquad \text{eq.2}$$

It is seen from Figure 4a, b that the EMI SE of TPU/GNP nanocomposite and microcellular foams increases in the investigated frequency range. Neat TPU has been found nearly transparent to electromagnetic wave with SE values lower than 2 dB; while, with increasing GNP content, the EMI SE of the nanocomposites increased gradually up to ~18 dB (at 1 THz) at 1 wt % filler loading, which is very close to the target value of EMI SE required for practical application (~20 dB). We have observed that even after the volume expansion, the EMI properties kept constant in terms of the total shielding. According to the approach proposed by Gupta et al.[5], the specific EMI shielding efficiency (EMI shielding efficiency divided by density, *SSE*) would be more appropriate when dealing with foamed shielding materials. Owing to the lower density of the foamed sheets, the SSE would be higher than the solid sheets[4]. Actually, the *SSE* for the TPU/GNP microcellular foam with higher filler content is 22.5 dB cm$^3$ g$^{-1}$, which is ~1.5 times higher than the non-foamed counterpart.

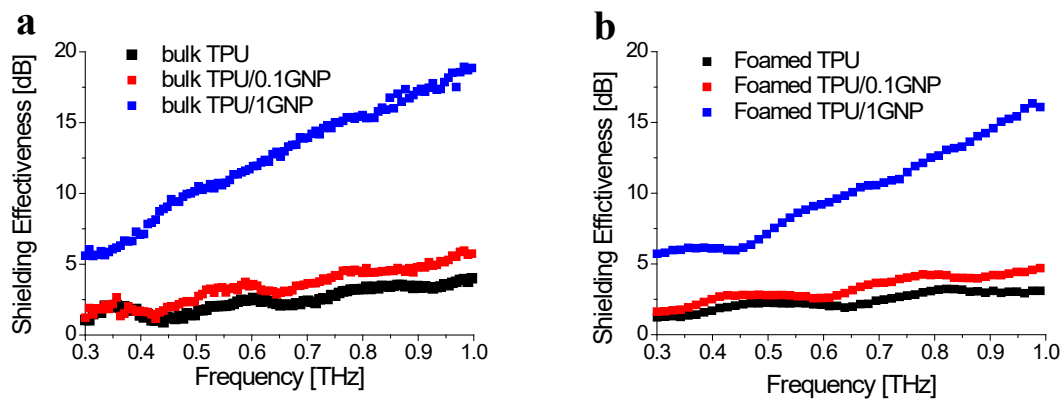

**Figure 4**. EMI shielding efficiency of TPU/GNP nanocomposites (a) and microcellular foams (b) at different frequency.

As EMI SE generally increases with increasing the specimen thickness, it is reasonable to expect that the EMI SE value for the TPU/GNP foams can be improved by increasing the specimen thickness and the graphene content. Therefore, if we divide the EMI SE by the density and the specimen thickness, thus obtaining the absolute shielding effectiveness of the material ($SSE_t$, measured in dB cm$^2$ g$^{-1}$), we can compare the performance of the produced microcellular foams with other materials. It is interesting noting that the $SSE_t$ estimated for the TPU/GNP microcellular foam with is relatively high in comparison with other polymeric foams filled with graphene-related materials presented in literature, which present similar specific $SSE_t$ values but for higher content of filler (10-30%)[17, 31-33].

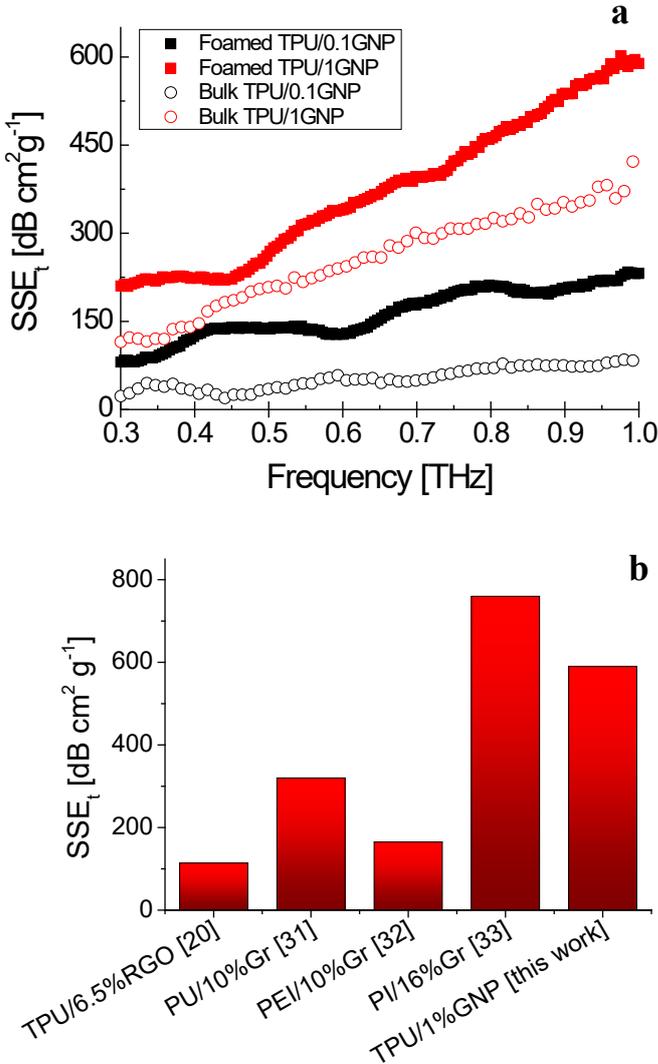

**Figure 5**. Absolute shielding effectiveness of the TPU/GNP microcellular foam as function of frequency (a) and comparative plot with other reported results on polymeric foams filled

with GRMs (b).

Furthermore, from the transfer function $\tilde{T}$ one can easily extract the frequency-dependent complex dielectric function[34]

$$\tilde{\varepsilon}(\omega) = \varepsilon'(\omega) + \varepsilon''(\omega) \qquad \text{eq.3}$$

that can be used to estimate the THz conductivity, given the relation

$$\sigma'(\omega) = \varepsilon_0 \omega \varepsilon''(\omega) \qquad \text{eq.4}$$

where $\varepsilon_o$ is the vacuum permittivity.

The retrieved σ' values together with the experimental $SSE_t$ data at 1 THz are shown in Table 2. As expected, conductivity goes up with the increase of GNP content for both the bulk and foamed nanocomposites.

| Sample | σ'[S/m] | SSE [dB cm² g⁻¹] |
|---|---|---|
| Foamed TPU/0.1GNP | 13 | $2.3 \cdot 10^2$ |
| Foamed TPU/1GNP | 41 | $6.0 \cdot 10^2$ |
| TPU/0.1GNP | 13 | $0.8 \cdot 10^2$ |
| TPU/1GNP | 51 | $3.7 \cdot 10^2$ |

**Table 2**. Conductivity and $SSE_t$ values for foamed and non-foamed TPU/GNP nanocomposites at 1THz.

## 3. Conclusions

GNP/TPU microcellular foams were prepared by solution blending and then foaming with an eco-friendly supercritical $CO_2$ foaming technique. The incorporation of only 1 wt% of GNP lead to EMI shielding efficiency of 16-18 dB, for the bulk and foamed nanocomposites, which is close to the target value of EMI SE required for practical application. Furthermore, the foamed TPU/GNP shows an absolute shielding efficiency of ~600 dB cm² g⁻¹, which is comparable to other polymeric foams reinforced with much higher amount of graphene-related materials.

**Acknowledgments**

This activity has received funding from the European Union's Horizon 2020 research and innovation programme under grant agreement No GrapheneCore3 881603. The authors acknowledge also the financial support of the National Institute for Nuclear Physics (INFN) under the project "TERA".